\documentclass[preprint,superscriptaddress]{revtex4}
\usepackage{graphicx}
\usepackage{array}
\usepackage{amssymb}
\usepackage{amsfonts}
\usepackage{amsmath}
\usepackage{mathrsfs}
\usepackage{color}
\usepackage{booktabs}
\usepackage{threeparttable}
\usepackage{multirow}
\usepackage{subfigure}
\usepackage{times}
\usepackage{epsfig}
\usepackage{chngpage}
\usepackage{latexsym}
\usepackage{mathrsfs}
\usepackage{bm}
\usepackage{epstopdf}
\begin{document}

\title{Locating the source of diffusion in complex networks by time-reversal backward spreading}

\author{Zhesi Shen}
\affiliation{School of Systems Science, Beijing Normal University, Beijing, 100875, P. R. China}

\author{Shinan Cao}\email{shinancao@ruc.edu.cn}
\affiliation{School of Finance, University of International Business and Economics, Beijing, 100029, P. R. China}

\author{Wen-Xu Wang}\email{wenxuwang@bnu.edu.cn}
\affiliation{School of Systems Science, Beijing Normal University, Beijing, 100875, P. R. China}
\affiliation{Business School, University of Shanghai for Science and Technology, Shanghai 200093, China}

\author{Zengru Di}
\affiliation{School of Systems Science, Beijing Normal University, Beijing, 100875, P. R. China}

\author{H. Eugene Stanley}
\affiliation{Department of Physics, Boston University, Boston,
Massachusetts 02215, USA}

\begin{abstract}
Locating the source that triggers a dynamical process is a fundamental
but challenging problem in complex networks, ranging from epidemic
spreading in society and on the Internet to cancer metastasis in the
human body. An accurate localization of the source is inherently limited
by our ability to simultaneously access the information of all nodes in
a large-scale complex network. This thus raises two critical
questions: how do we locate the source from incomplete information and
can we achieve full localization of sources at any possible location
from a given set of observable nodes. Here we develop a time-reversal backward spreading algorithm to locate the source of a diffusion-like process efficiently and propose a general locatability condition. We test the algorithm by employing epidemic
spreading and consensus dynamics as typical dynamical processes and
apply it to the H1N1 pandemic in China. We find that the sources can be
precisely located in arbitrary networks insofar as the locatability
condition is assured. Our tools greatly improve our ability to locate
the source of diffusion in complex networks based on limited
accessibility of nodal information. Moreover, they have implications for
controlling a variety of dynamical processes taking place on complex
networks, such as inhibiting epidemics, slowing the spread of rumors,
pollution control and environmental protection.

\end{abstract}

\maketitle

{\it Introduction.-}Many large-scale dynamical processes taking place on complex
networks can be triggered from a small number of nodes. Prototypical
examples include epidemic spreading on a global scale, rumor propagation
through micro-blogs on the Internet, wide-ranging blackouts across North America
and financial crises accompanied by the bankruptcy of a large number of
financial institutions. The self-organization theory introduced by Bak
and his collaborators~\cite{bak1997nature} has provided a theoretical
explanation: when a complex system enters a self-organized criticality
state, small perturbations to even single individuals are able to
initiate a big event, such as the avalanche of collapses in the sandpile
model~\cite{tang-prl}. Moreover, the development of modern technology
considerably facilitates the spreading of disease and information via
public traffic systems and the Internet, which enables propagation
across a large area from a source, such as the worldwide H1N1 pandemic
in 2009~\cite{fraser2009pandemic,neumann2009emergence} and the
irrational and panicked acquisition of salt in southeast Asian countries
caused by a rumor relevant to the nuclear leak in
Japan. These phenomena raise a challenging question: how to
locate the source in a huge network relying on relatively limited
accessibility to nodal states, answers to which are of paramount
importance for many aspects of nature and society, such as disease
control, anti-terrorism, and economic health. Despite some
pioneering approaches attempting to locate sources~\cite{source-prl,brockmann2013hidden,source-BP,lokhov2013inferring,zhu2013information, zhu2013information2,antulov2015identification}
and superspreaders~\cite{superspreader1,superspreader2}, we still lack a
comprehensive understanding of our ability to precisely identify the
original source of spreading in a large complex network. The difficulty
stems from the lack of a general
locatability condition to predict if the source at any possible
locations is fully locatable in terms of a given set of observers.

We develop a general locatability framework based on the time
reversible characteristic of diffusion-like processes. This allows us to
perform a time-reversal backward spreading to accurately locate the source,
and offer a locatability condition that guarantees that a source will be
fully locatable at any position.
The algorithm and locatability condition are
applicable in both directed and undirected networks with inherently
limited knowledge of nodes and a time delay along links. We validate the
tools by using a variety of complex networks in combination with two
typical diffusion-like dynamical processes, i.e., epidemic
spreading~\cite{anderson1991infectious,colizza2007invasion,balcan2009multiscale}
and consensus dynamics~\cite{consensus-1,olfati2007consensus}. We have also applied
our method to real networked systems by employing empirical data from
the 2009 H1N1 pandemic in China, focusing on the Chinese airline and
train networks as the epidemic spreading network. The four sources
predicted by our tools are in good agreement with empirical
findings. Our framework has further potential applications in locating,
for example, a spammer who abuses email systems and pollution sources in river networks.



{\it Time-reversal backward spreading.-} Our goal is to locate the source that initiates a diffusion-like process
taking place on an already-known undirected or directed complex network
using only the limited time information pertaining to the diffusion
observed from a fraction of nodes. This limited information could be the
time period during which a person is being invaded by a virus, or the
appearance of an abnormal signal at a node. To better mimic a
real-world scenario, we assume that we are unable to detect
communications between the observable nodes and their neighbors.
For example, hospital records tell us when a
patient became ill, but do not tell us who passed the disease to the
patient. Even knowing all of the sick persons with whom the patient has
had recent contact does not tell us.

\begin{figure*}[htb]
\begin{center}
\epsfig{figure=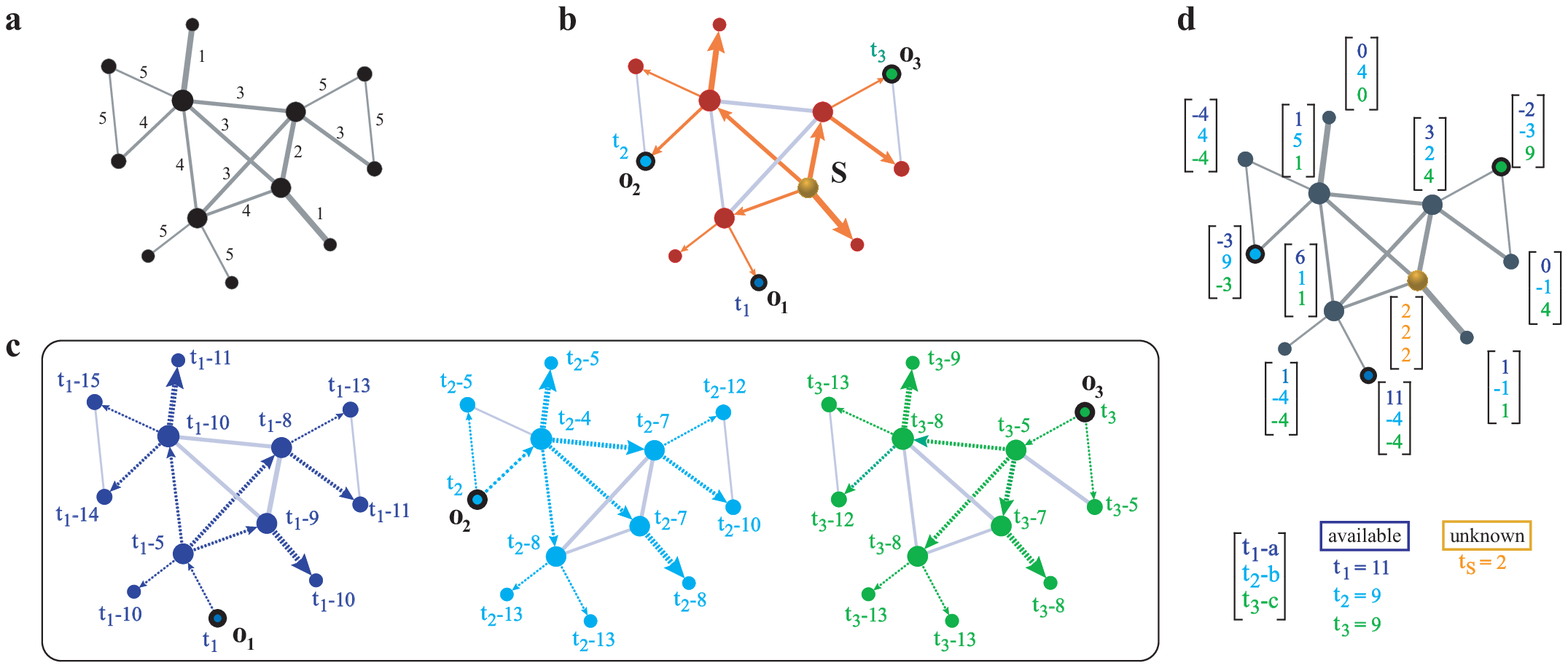,width=\linewidth} 
\caption{{\bf Time-reversal backward spreading for locating the source.} {\bf a}, a
network topology with link weights (time delay). {\bf b}, the diffusion paths from the source $S$ and the observers $o_1$, $o_2$ and $o_3$. The arrival time only at the three observers, namely, $t_1$, $t_2$ and $t_3$ can be accessed. {\bf c}, implement TRBS along weighted shortest paths from $o_1$, $o_2$ and $o_3$, respectively and the reversed arrival time at each node stems from each observer, respectively. {\bf d}, the vector $\mathbf{T}$ consisting of the reversed arrival time from each of the observers. The elements of $\mathbf{T}_s$ of the source are identical, which is the key to distinguishing the source from the other nodes. If the observers provide sufficient information of the source, the revered arrival time from observers are the original time $t_s$ of the diffusion from the source, enabling the recovery of $t_s$. The source $S$ is in yellow and the three observer nodes are in dark blue, light blue and green with black boundary. The actual diffusion from $S$ is marked by orange solid lines with arrows and the TRBS from the observers are respectively marked by colored dotted lines with arrows. The color of numbers in the vector in (d) corresponds to the observer with the same color.  }
\label{fig:algorithm}
\end{center}
\end{figure*}

The network and the spreading process are illustrated in Figs.~\ref{fig:algorithm}(a) and (b), respectively. The weights along links are the time delay of passing a signal along links. For an undirected network, the delay along a link is the same for both directions. Figure~\ref{fig:algorithm}(b) shows that a spreading process starts from source node $s$ and propagates from the source to the whole network along the weighted shortest pathes to all nodes (because the shortest pathes are associated with the shortest propagation delay).

Our time-reversal backward spreading (TRBS) algorithm for locating sources is based solely on (i) the weighted network structure (Fig.~\ref{fig:algorithm}(a))
and (ii) the arrival time of certain signals at nodes
that we call observers. These accessible observers $o_1, o_2,\cdots,o_m$
receive a signal at time $t_{\text{o}_1},t_{\text{o}_2},\cdots,
t_{\text{o}_m}$, as shown in Fig.~\ref{fig:algorithm}(b). We assume the source $s$, the original time $t_s$ at $s$, and the diffusion pathes from $s$ are
unknown. Because of the stochastic effect in real-world networked
systems, we may not know the exact propagation delay along a link
between two nodes, but we assume that the time delay follows a certain
distribution, e.g., the Gaussian or uniform distributions. Insofar as
the mean value and variance are finite, which are commonly observed in
real scenario, our algorithm is feasible if we use the mean delay. If
the distributions of time delay on ecah link are nonidentical, we can use the mean
value of each link to specify the time delay of each link.
The TRBS algorithm based on
the weighted network and the signal arrival time at some observers
is as follows:

\begin{itemize}

\item[{(i)}] Perform the TRBS starting from an observer
  $o_k$ to all nodes in the networks along the reversed direction of links
   (for a directed network, TRBS from node $i$ to $j$ is allowed if and only if there is a directed link with direction from $j$ to $i$, namely the reversed direction of the link; for an undirected network, links are bidirectional with the same time delay on both directions and the reversed direction is the same as the original direction). This yields a reversed arrival time $t_{o_k} -
  \hat{t}(i,o_k)$ at an arbitrary node $i$, where $\hat{t}(i,o_k)$ is
  the shortest time delay from $o_k$ to $i$ (see
  Fig.~\ref{fig:algorithm}(c)). Thus the set of observers leads to a
  vector $\mathbf{T}_i=
  [t_{\text{o}_1}-\hat{t}(i,o_1),t_{o_1}-\hat{t}(i,o_2),\cdots,t_{\text{o}_1}-\hat{t}(i,o_m)]^\text{T}$
  for node $i$ (see Fig.~\ref{fig:algorithm}(d)). Note that the reversed arrival time is a virtual time for source localization.

\item[{(ii)}] Calculate the variance of the elements in
  $\mathbf{T}_1,\mathbf{T}_2,\cdots,\mathbf{T}_N$. The node with the
  minimum variance is the source (see Fig.~\ref{fig:algorithm}(d)). Using
  our algorithm we can locate the source with computational amount $O(mN\log{N})$,
  and $O(N^2 \log{N})$ in the worse case,
  where $m$ is the number of observers, $N$ is the number of nodes, and
  $m<N$.

\end{itemize}

For an idealized scenario in which we know the exact time delay (weight) along
each link, the source will have zero variance (see
Fig.~\ref{fig:algorithm}(d)). Since the diffusion process is reversible,
the time-reversal delay from $o_k$ to $s$ is equal to the actual delay
from $s$ to $o_k$, i.e., $t_{o_k}-t_s = \hat{t}(s,o_k)$, which leads to
$t_{\text{o}_1}-\hat{t}(s,o_1)=
t_{\text{o}_2}-\hat{t}(s,o_2)=\cdots=t_{\text{o}_m}-\hat{t}(s,o_m)=t_s$
with zero variance. In contrast, for a node other than $s$ the paths of TRBS from the observers will not be the same as that of the
actual paths of spreading from the source, and node variance will be nonzero.

{\it Locatability condition.-} We offer a locatability condition to determine if a source
at any possible location can be fully localized from the arrival time $t_{o_k}$
($k=1,\cdots,m$) at arbitrary $m$ given observers. Based on the vector
$\mathbf{T}_i$ ($i=1,\cdots,N$) calculated from $m$
observers, we define the difference between the vector of any two nodes
$i$ and $j$, $\Delta \mathbf{T}_{ij} \equiv \mathbf{T_i}-\mathbf{T_j}$.
The locatability condition can then be given: if and only if the
elements of $\Delta \mathbf{T}_{ij}$ for any two nodes are not all the
same, the source at any location can be exactly identified.

The general locatability condition is equivalent to the statement that if there exist any two nodes, say, $i$ and $j$, such that the elements of their $\Delta \mathbf{T}_{ij}$ are the same, the source cannot be distinguished between $i$ and $j$. In the following, we justify this equivalent locatability condition. We first describe the equivalent condition mathematically. Let's denote the shortest time delay from node $i$ to observer
$o_k$ by $\hat{t}(i,o_k)$ that is defined as
\begin{equation}
\hat{t}(i,o_k) = \sum_{\nu \in P(i,o_k)} \theta_\nu,
\end{equation}
where $\theta_\nu$ is the time delay along link $\nu$ and $P(i,o_k)$ denotes
the set of shortest weighted path between $i$ and $o_k$.
Since the diffusion process is reversible along reversed links, 
according to the definition of $\mathbf{T}_i$, we have
\begin{eqnarray}
\Delta \mathbf{T}_{ij} &=& \mathbf{T}_{i} - \mathbf{T}_j =\left(
                                                            \begin{array}{c}
                                                            \hat{t}(j,o_1)-\hat{t}(i,o_1) \\
                                                            \hat{t}(j,o_2)-\hat{t}(i,o_2) \\
                                                            \vdots \\
                                                            \hat{t}(j,o_m)-\hat{t}(i,o_m)
                                                            \end{array}
                                                          \right).
\end{eqnarray}
If the locatability condition is violated, namely,
\begin{eqnarray}
\hat{t}(j,o_1)-\hat{t}(i,o_1) &=& \hat{t}(j,o_2)-\hat{t}(i,o_2) = \cdots  \nonumber \\
&=&\hat{t}(j,o_m)-\hat{t}(i,o_m),
\label{eq:allthesame}
\end{eqnarray}
we cannot identify the source $s$ when $s\in (i,j)$, which is the equivalent locatability condition and can be proved as follows. Assume that $i$ is the actual source with original time
$t_i^s$ and node $i$ and $j$ satisfies Eq.~(\ref{eq:allthesame}). The
source $i$ gives rise to the arrival time
$t_{\text{o}_1},t_{\text{o}_2},\cdots,t_{\text{o}_m}$ at observers
$o_1,o_2,\cdots,o_m$. Suppose that $j$ is the source and the original time at $j$ is $t_j^s$, which leads to the arrival time
$t'_{\text{o}_1},t'_{\text{o}_2},\cdots,t'_{\text{o}_m}$ at the same set
of $m$ observers (for the source, origin time is the same as arrival time).
Taking the time reversible characteristics of TRBS along reversed links, we can simply have $t_{o_m}=\hat{t}(i,o_m)$ and $t'_{o_m}=\hat{t}(j,o_m)$. According to Eq.~(\ref{eq:allthesame}), we can derive that $t_{\text{o}_1}-t'_{\text{o}_1}=t_{\text{o}_2}-t'_{\text{o}_2}=\cdots=t_{\text{o}_m}-t'_{\text{o}_m}=t_i^s-t_j^s+c$,
where $c$ is a constant. Note that if the original time at $j$ is $t_j^s = t_i^s + c$, we have $t_{\text{o}_1}-t'_{\text{o}_1}=t_{\text{o}_2}-t'_{\text{o}_2}=\cdots=t_{\text{o}_m}-t'_{\text{o}_m}=t_i^s-t_j^s+c = 0$, which indicates that source $i$ and source $j$ generate exactly the same arrival time as the actual observed arrival time at all the observers. Thus, the source cannot be distinguished between $i$ and $j$ in principle.
In other words, because the actual original time $t_s$ is unknown, if Eq.~(\ref{eq:allthesame}) is satisfied, there exists two possible original time $t_i^s$ and $t_j^s$ with $t_j^s=t_i^s+c$, such that the spreading process starts from node $i$ and $j$ will generate the same arrival time as the actual arrival time at observers, rendering the source between $i$ and $j$ indistinguishable. Hence, our locatability condition offers a sufficient and necessary criterion for exclusively locating the source. If the locatability condition is satisfied, namely, Eq.~(\ref{eq:allthesame}) is violated, at least one observer is
able to provide effective information that is sufficient to distinguish
$i$ and $j$ by using, for example, our efficient algorithm. Therefore,
the source in a network is said locatable if and only if for any two
nodes $i$ and $j$, the element values in $\Delta \mathbf{T}_{ij}$ are
not all the same.

\begin{figure}[htb]
\begin{center}
\epsfig{figure=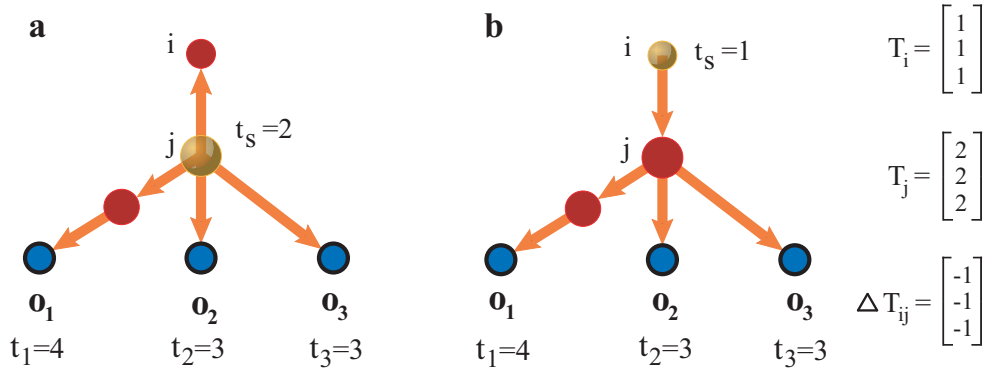,width=0.8\linewidth}
\caption{{\bf The uncertainty of source.} {\bf a}, a diffusion process from the
  source $j$ at $t_s=2$ with three observers $o_1$, $o_2$ and
  $o_3$. {\bf b}, a diffusion from the source $i$ at $t_s=1$ with the
  same observers as in ({\bf a}). The source in ({\bf a}) and ({\bf b})
  produces the same arrival time at the three observers, i.e., $t_1$, $t_2$ and $t_3$. {\bf c}, the
  vector $\mathbf{T}_i$ and $\mathbf{T}_j$ and the difference $\Delta
  \mathbf{T_{ij}}$ between them. Without loss of generality, we assume
  the time delay along each link is 1. The original time $t_s$ of the
  diffusion from a source is known for the locatability problem. The
  color of nodes and links represents the same meaning as that in
  Fig.~\ref{fig:algorithm}.  }
\label{fig:locatability}
\end{center}
\end{figure}

Figure~\ref{fig:locatability} gives an intuitive example to explain the locatability condition. Since the original time $t_s$ at the source is unknown, if we choose a certain original time, e.g., $t_s=1$ at node $i$ or $t_s=2$ at node $j$, both nodes can produce the exact same arrival time at the
three observers ($t_1=4$, $t_2=3$ and $t_3=3$),
indicating that the source cannot be distinguished between $i$ and $j$.
Thus the source in the network with respect to the
given set of observers is not locatable. This scenario is exactly reflected by
$\Delta \mathbf{T}_{ij}$ in which all elements are the same. The
locatability condition in principle inhibits the indistinguishable
scenario and exclusively locating the source at any location is assured. If the locatability condition is satisfied, namely, there
is a single node in which the elements in its vector $\mathbf{T}_\text{s}$
are identical, this identical value is the original time of the diffusion from the
source. This is because of the intrinsic time-reversal characteristic of the TRBS process. When implementing the TRBS, the reversed arrival time at the source is nothing but the original time $t_s$ that is the identical value in the vector $\mathbf{T}_\text{s}$ of the source, as shown in Fig.~\ref{fig:algorithm}(d).
Therefore, if the source in a complex network is fully locatable, the original time
of diffusion can be inferred as well.

An immediate consequence of the locatability condition is that a node
with a single neighbor must be observed to guarantee fully locatable.
This can be easily proved by noting that the node and any one of its
neighbors cannot be distinguished for any observers, except the node
itself according to Eq.~(\ref{eq:allthesame}). This consequence
indicates that for a star graph, all nodes except the star should be
observed, and in a tree, we usually need to observe a large fraction of
nodes to enable full localization. For a fully connected network with
$N$ nodes, we must observe $N-1$ nodes to assure fully locatable.
For an undirected chain, both ends should be observed for
locating a source.



Note that the locatability condition is rigorous for idealized networks
in which we know the exact time delay along each link. In practice, if
the time delay of a link follows some distribution resulting from the
stochastic effect, the locatability condition is violated somewhat. This
is analogous to the structural observability~\cite{stc} of those scenarios in which
we lack a complete knowledge of link weights. Despite this lack, it is
possible for us to use the locatability condition to identify a source
from a pair of nodes. If the element values of $\Delta \mathbf{T}_{ij}$
are sufficiently close, it is likely that nodes $i$ and $j$ will be
indistinguishable. If the element values differ greatly, however, it is
easier for us to identify which one is more likely to be the source between them.

{\it Source localization performance.-} To validate our locatability framework we explore two prototypical
dynamical processes, diffusion and consensus. Diffusion processes commonly
occur in many natural and social network systems, such as epidemic spreading
in a population, virus propagation on the
Internet~\cite{lloyd2001viruses,kleinberg2007epidemic}, rumor
propagation in social networks~\cite{zanette2002dynamics}, and risk
contagion in financial networks~\cite{gai2010contagion}. Some dynamical
processes are not subject to diffusion, but exhibit diffusion-like
behavior, e.g., cascading failures in power
grids~\cite{motter2002cascade,buldyrev2010catastrophic,schneider2011mitigation,gao2011networks}
and the spreading of gridlock in urban automobile traffic
patterns~\cite{duch2006scaling,wu2008traffic,Chaos-Yang2012}. To be as general as
possible, we consider the simplest diffusion model, the one associated
with diffusion delay. To simulate a diffusion process, we must first
construct a complex network with a node degree distribution that allows
the diffusion of a signal, e.g., a virus, a rumor, or a risky behavior
in social network. Each link is assigned a time delay (weight) of forwarding the signal and the weights of links can be either the same or follow a distribution. The simulation is carried out as
follows. First, a randomly selected source passes the signal to its
neighbors. The signal takes some time to reach its neighbor nodes,
depending on the link delays. Each node that has received the signal
forwards it to its neighbors and this process continues until all the
nodes in the network have received the signal. What we can measure and
record is the arrival time of the signal at the observer nodes.


Consensus dynamics on complex networks have been investigated since the
development of complex network science a decade
ago~\cite{newman2009networks,lynch1996distributed,olfati2006flocking,vicsek2012physicreport,
vicsek2013pnas,egerstedt2001formation,savkin2004coordinated}.
Although most real systems display nonlinear
behavior, agreement and synchronization phenomena are in many aspects
similar to the consensus of linear systems. We thus use simple canonical
linear, time-invariant dynamics with a communication
delay~\cite{olfati2007consensus}
\begin{equation}
\dot{x}_i = \sum_{j=1}^{N} a_{ij}[x_j(t-\tau_{ij}) - x_i(t)],
\label{eq:consensus}
\end{equation}
where $x_i(t)$ $(i=1,\cdots,N)$ is the state of node $i$ at time $t$,
and $\tau_{ij}$ is the time delay along the link between node $i$ and
node $j$. We explore the
diffusion of a perturbation starting from a single source node in the consensus
state. Note that, unlike the standard diffusion process via contact or
transportation, the diffusion-like process of perturbation is caused by
the node coupling. Specifically, in the absence of external
perturbations, all nodes uniformly stay in the consensus state. Thus
the transmission of a signal to other nodes can be discerned when
deviation from the consensus state occurs. We record the time at which
the state of observable nodes deviates from the consensus state and, using our
locatability framework to locate the source node with original perturbation.

\begin{figure}[htb]
\begin{center}
\epsfig{figure=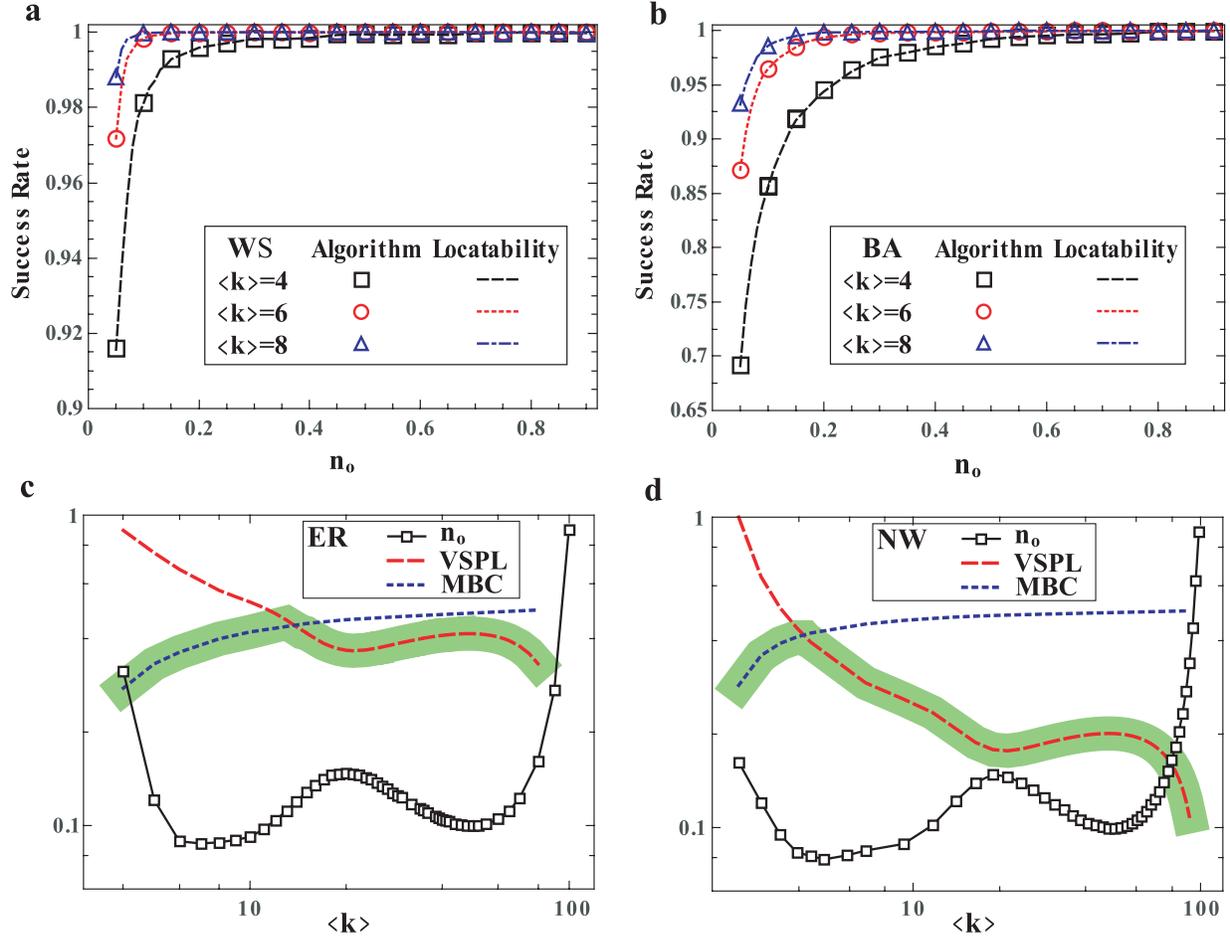,width=\linewidth}
\caption{{\bf Locatability of source in model networks.} {\bf a}-{\bf
    b}, success rate obtained using the efficient algorithm and
  predicted by the locatability condition in Watts-Strogatz (WS)
  small-world network ({\bf a}) and Barab\'asi-Albert (BA) network ({\bf
    b}) for different average node degree $\langle k\rangle$. {\bf
    c}-{\bf d}, the minimum number $n_\text{o}$ of observers to reach
  $90\%$ success rate, the effect of the maximum betweenness centrality
  (MBC) and the variance of shortest path length (VSPL) as a function of
  $\langle k\rangle$ respectively in Erd\"os-R\'enyi (ER) random network
  ({\bf c}) and Newman-Watts (NW) small-world network. The green belt
  represents the joint effect of MBC and VSPL on the locatability.
  The numerical results are obtained by averaging over
  400 independent realizations and the network size is 100.  }
\label{fig:simu}
\end{center}
\end{figure}

We numerically validate our locatability condition by comparing with the success rate of locating sources when the exact weights of links are known. Figure~\ref{fig:simu}(a) and (b) show the success
rate of locating sources in small-world and scale-free networks by using our TRBS algorithm. It shows exact agreement with the prediction of the
locatability condition for both homogeneous and inhomogeneous networks
with a different average node degree $\langle k \rangle$ and fraction of observers $n_o$. The success rate achieves
the upper bound predicted by the locatability condition, indicating that
our TRBS algorithm is optimal for locating the source of spreading.
Figures~\ref{fig:simu}(c) and \ref{fig:simu}(d) show the minimum
fraction $n_\text{o}^{\min}$ of randomly-chosen observers to reach $90\%$ success rate affected by $\langle k\rangle$ in random and small-world networks. Note that
$n_\text{o}^{\min}$ exhibits a w-shape function of $\langle k\rangle$
with two optimal values of $\langle k\rangle$. This counterintuitive
finding can be understood in terms of the change of the maximum
betweenness centrality (MBC) and the variance of the shortest path
length (SPL). Their joint effects on $n_\text{o}$ can be heuristically
explained based on the locatability condition. On the one hand, let's consider a scenario that node $i$ must be passed in order to reach node $j$ along the shortest path
from the observers. In this case, the source between $i$ and $j$ will be indistinguishable (see Fig.~\ref{fig:locatability}). If this occurs, the number of the observers is approximately equal to the betweenness centrality of $i$. Hence highest the probability of encountering this scenario for any two nodes is
reflected in the MBC in the network. The larger MBC means
that there is a higher probability that the locatability condition will
be violated, and this accounts for the requirement of more observers, namely, the higher value of $n_\text{o}$. On
the other hand, $n_\text{o}$ is affected by the variance of the shortest
path length in the network. If the shortest paths from all the
observers to node $i$ and $j$ are the same, based on the locatability
condition, the source between $i$ and $j$ will be indistinguishable in the sense that the reversed arrival time at both nodes are exactly the same. An extreme case is the fully-connected network with zero variance of SPL in which $N-1$
observers are needed. Thus a larger variance of SPL results in lower
values of $n_\text{o}$. The joint effect of BC and SP on $n_\text{o}$
gives rise to the ``w-shape'' with two optimal average degrees, as shown
in the green region in Figs.~\ref{fig:simu}(c) and \ref{fig:simu}(d).

\begin{figure}[htb]
\begin{center}
\epsfig{figure=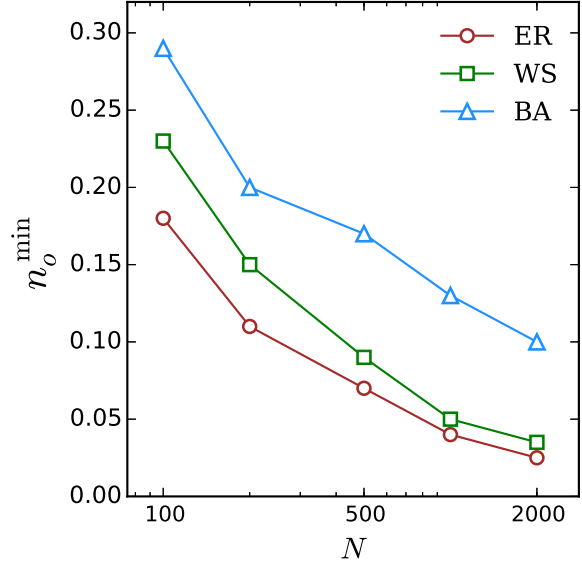,width=0.5\linewidth}
\caption{{\bf Minimum fraction of observers for different network size.} The minimum fraction $n_\text{o}^{\min}$ of randomly selected observers that assures $90\%$ success rate of locating the source of spreading process on ER, WS and BA networks. The time delays of links are assumed to follow Gaussian distributions with mean value 1.0 and variance 0.25. The average node degree $\langle k\rangle =8$. The results are obtained by averaging over 500 independent realizations.}
\label{fig:size}
\end{center}
\end{figure}

Table~\ref{tab:simu} displays $n_\text{o}^{\min}$ for achieving a 90\%
success rate of locating the source in homogeneous and heterogeneous
networks associated with a Gaussian distribution and a uniform distribution
of time delay along links, respectively. We assume that only the mean time delay along links rather than the exact time delay along each links is known. We assign the mean time delay to each link, such that the network becomes a weighted network with identical link weights. The results
demonstrate that our algorithm is successful based on the mean time delay without exact time delays along links for both spreading and consensus dynamics. {The small differences between $n_\text{o}^{\min}$ of spreading process and consensus dynamics are resulting from the approximation during the numerical integral of Eq.~\ref{eq:consensus}. Figure~\ref{fig:size} shows the relations between $n_\text{o}^{\min}$ and network size $N$. As we can see, the fraction of required observers decreases as the network size increases for all the model networks, implying the effectiveness and applicability of our method. We also compares the performance with the Jordan Center method~\cite{zhu2013information2}, which is an topology based method, shown in Table~\ref{tab:jordan}. The average rankings of the real source node in our algorithm approaches 1, which is much smaller than the rankings in Jordan center method.} The robustness of our method under conditions of incomplete information and noisy data, and its need
for only a small fraction of observers allows it to be generally
applicable in real-world networked systems in which conditions of
measurement noise and incomplete node information are inevitable.

{\it Locating the source of H1N1 spreading in China.-}
We apply our locatability framework to the H1N1 pandemic in China in
2009. We use the empirical data to quantify the arrival time of the
virus at each major city to discern the source with the earliest arrival time. Note that
we assume that only the arrival time of a fraction of major cities are accessible and we aim to locate the source from the arrival time.
We use both airline and train networks among provinces to capture the spreading network{, in which the total number of vertex is 31}. The airports and train stations are usually located
at the provincial capital cites, and the bidirectional links between
two nodes are weighted and related with the customer flux estimated
by the number of flights and trains per day. The time delay $\tau$ along each link is estimated from the flux of passengers in unit time
by the following formula
\begin{equation}
\tau_{ij}=\frac{1}{1-(1-\varphi)(1-\xi)^{w_{ij}}},
\label{eq:timedelay}
\end{equation}
where $i$ and $j$ represent two major cities,
$\varphi$ characterizes the time scale of the spreading process,
$\xi$ is the probability of a single infected passage taking an airplane
or a train, $w_{ij}$ is the number of equivalent airplanes per day
between $i$ and $j$. $w_{ij}$ is set according to China airline and train
data base, where a train is equivalent to 5 airplanes. $\varphi$ is set
to be $1/4$, due to the fact that H1N1 pandemic in China last for about
4 months with the time unit 1 month. $\xi$ is fixed to be $1/2000$ owing
to the fact that on average there are about 300 available seats per
airplane and about 1600 available seats per train with the sum is about
2000. We have checked that our results of locating the source is
insensitive to the value of $\xi$. In the range of $1/1800 <\xi <
1/3000$, our algorithm offers approximately the same locating
probability of the source. The dominator of
Eq.~(\ref{eq:timedelay}) captures the infection probability between $i$
and $j$, so that the reciprocal of the infection probability corresponds
to the time delay.

\begin{figure*}[htb]
\begin{center}
\epsfig{figure=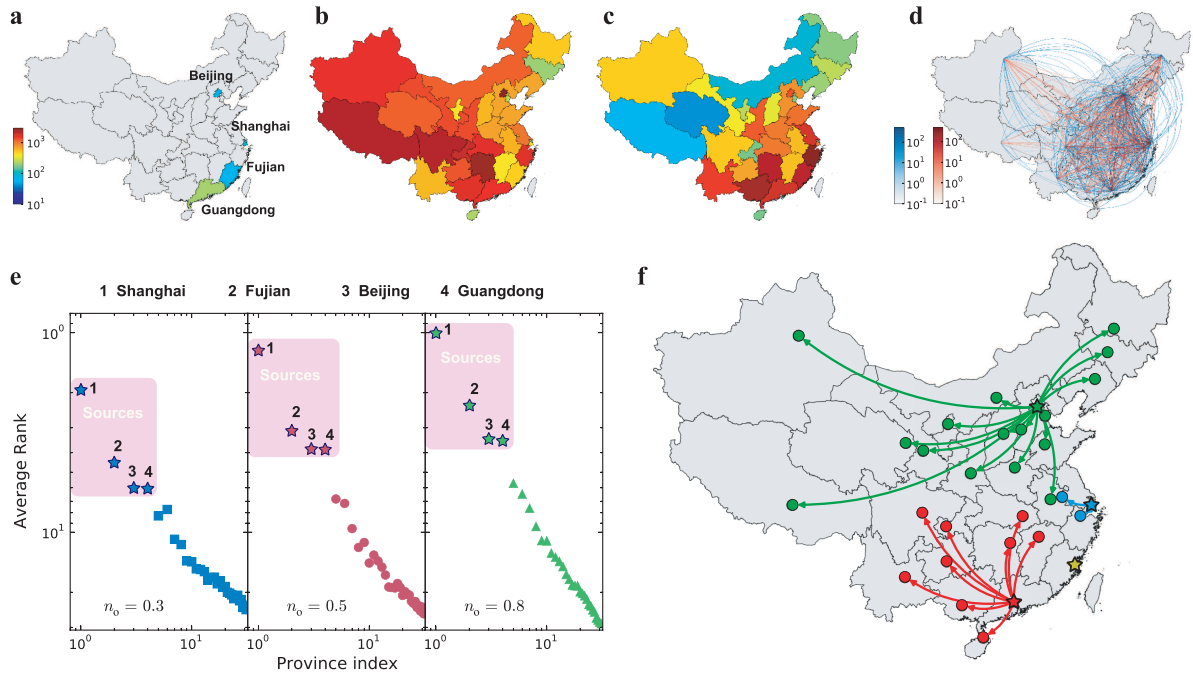,width=\linewidth}
\caption{{\bf Locate the sources of H1N1 pandemic in China.} {\bf a},
the earliest outbreak of H1N1 in June 2009 in four
provinces---Beijing, Shanghai, Fujian and Guangdong---which are the
sources of the epidemic spreading in China.  The epidemic outbreaks
occur at the four locations nearly simultaneously. {\bf b}, the
outbreak of H1N1 all over China in Oct. 2009. {\bf c}, The number of
patients in China in Dec. 2009. The color bar in ({\bf a}), ({\bf b})
and ({\bf c}) denote the number of patents. {\bf d}, China airline and
train networks with weighted links. The color bars capture the
passenger flux of airlines and trains per day, respectively. The
mixture of the airline and train networks is used as the propagation
network of the H1N1 virus. {\bf e}, the average ranks of different provinces
corresponding to the probabilities of being the sources of the
epidemic spreading calculated by our algorithm. The four actual
sources are of the highest four ranks with respect to different
fraction $n_\text{o}$ of observers and there is a clear gap between
the sources and the other provinces. {\bf f}, the most probable paths
of spreading from the sources uncovered by using the estimated time
delays along links. The results in ({\bf e}) are obtained by randomly
choosing 100 independent configurations of observers with different
fractions.  }
\label{fig:H1N1}
\end{center}
\end{figure*}

Figure~\ref{fig:H1N1} (a) to (c) show the empirical record of the H1N1 pandemic in China in 2009. Specifically,
Fig.~\ref{fig:H1N1}(a) shows that the
disease arises almost simultaneously from Beijing, Shanghai, Fujian, and
Guangdong, i.e., these four provinces are the
sources. Figure~\ref{fig:H1N1}(b) shows the outbreak of the disease
across China. Figure~\ref{fig:H1N1}(c) shows the
application of medical treatment after the epidemic has spread across
the country causes the number of cases to decrease and, some months
later, disappear. Figure~\ref{fig:H1N1}(d) shows both airline and train
networks in China with different passenger fluxes along the links. We
randomly pick a fraction of nodes to be observers and record
the outbreak time in each of them to be the arrival time, and use the combined network of flight and train to locate the disease sources (each province is a node with location represented by the major city in the province). In particular, for a
group of observers, we rank all the provinces according to their
probability of being a source as revealed by the variance of the elements in their
reversed arrival time vector ${\bf T}_i$. A node with smaller variance in ${\bf T}_i$ will be of higher probability to be a source.
Figure~\ref{fig:H1N1}(e) shows that the four nodes are found
to have the highest average ranks by the independent realizations for
different fractions of observers. Note that for $n_\text{o}>0.3$, there
is a clear gap between the average rank of the four provinces and that of the
other provinces, indicating the presence of four sources. As
$n_\text{o}$ increases, the gap widens, which is a strong evidence that
multiple sources exist. The four sources identified by our method are in exact agreement with the empirical record in Fig.~\ref{fig:H1N1}(a), validating the practical applicability of our method.
From the locations of the sources the most
probable spreading paths of the disease can be ascertained based on the
estimated time delay, as shown in Fig.~\ref{fig:H1N1}(f). The spreading
paths are obtained by preserving all paths with the shortest time delay
from one of the sources in the set of all infection paths. The hidden
radial spreading patterns from the sources are then uncovered using our
locatability framework.

The fact that the H1N1 virus came from outside China accounts for the
four sources that spurs the epidemic spreading in China. The four
source provinces have international airports and we suspect that the virus may
invade China via international flights from other countries. Despite the challenge
of more than one sources, our algorithm still offers quite high accuracy
of ascertaining all the sources, demonstrating the general applicability
of our approach for addressing real problems.

{\it Discussion and conclusion.-}
In a huge network often only a subset of nodes is accessible. We thus need an
efficient algorithm for locating the sources and ascertaining whether a
given set of observers provide sufficient information for source localization. Our locatability framework uses the time-reversal backward spreading process
on complex networks to provide tools to address these
fundamental questions. Our algorithm uses the arrival time of a signal
at the observers, the minimum information required, to locate the
source. Our general locatability condition also enables us to
determine whether the source in a network is fully locatable from a give set of observer nodes. We have systematically
tested our theoretical tools using diffusion processes and consensus
dynamics. Among the findings, an interesting result is the presence of two optimal
locatabilities as the link density increases from a very sparse network to a
fully-connected network. We have also applied our tools to H1N1
pandemic in China in 2009, finding that the four earliest-outbreak
provinces identified by our method from a small fraction of observers
are in good agreement with real data. Our theoretical tools have
implications for many dynamical processes pertaining to disease control,
identification of rare events in large networks, protection of the
normal functioning of the Internet, and the behavior of economic
systems.

{Our work still has some limitations. For example, the time delay along each link is assumed to be known, while, in many real situations, we can not get the time delays. How to accurate approximate the time delays with effective delays or equivalent delays, like the concept of effective distance in Ref~\cite{brockmann2013hidden}, when time delays are unavailable needs further investigation.}
In addition, our work raises a number of fundamental questions, answers to which
could further improve our ability to locate the source of diffusion-like dynamics occurring on complex networks. First, how
do we identify a minimum number of observers in an arbitrary network
using the locatability condition? Second, how
do we locate the sources using current methods if only part of the
network structure is accessible? We may overcome this obstacle by
using a network reconstruction approach based on the recently developed
compressive sensing method~\cite{cs-prl,cs-epl,cs-pd,cs-hidden}. Third,
how do we rank the observers with respect to the amount of effective
information they provide if the resources are limited and only a small
fraction of nodes are accessible? {Fourth, how to incorporate with the information of time delay variance and improve the performance if the whole time distribution is provided. The ideas in the Ref~\cite{antulov2015identification} may give some hints for better using the information of time delay variance.} Taken together,
our tools, because of their lower information requirements and solid
theoretical supports, could open new avenues for understanding and
controlling complex network systems, an extremely important goal in
contemporary science.

\begin{acknowledgments}

The authors thank Xiaoyong Yan, Xiao Han and Ying Fan for valuable discussions and help. This work was supported by NSFC under Grant Nos. 61174150, 61573064 and 61074116, the Fundamental Research Funds for the Central Universities and the Beijing Nova Programme.

\end{acknowledgments}

\begin{table}[htb]
\centering
\caption{{\bf Minimum fraction of observers.} The minimum fraction
      $n_\text{o}^{\min}$ of randomly selected observers that assures $90\%$
      success rate of locating the source of spreading process and the
      propagation of perturbation in consensus dynamics on ER, WS and BA
      networks. The time delays of links are assumed to follow Gaussian
      distributions with mean value 1.0 and variance 0.25 and uniform
      distributions in the range $(0.5,1.5)$, respectively. We exclusively
      use the mean delay of all links to identify sources. The network size
      $N$ is 100 and the average node degree $\langle k\rangle =8$. The
      results are obtained by averaging over 500 independent realizations.}
\begin{center}
\begin{threeparttable}
\begin{tabular}{cccc}
  \hline \hline \multirow{2}{*}{} & ER \ \ \ & WS \ \ \ & BA \ \ \ \\ &
  \multicolumn{3}{c}{(Gaussian / Uniform)}\\ \hline Spreading \ \ & 0.18
  / 0.23 \ \ & 0.23 / 0.36 \ \ & 0.29 / 0.41 \\ Consensus \ \ & 0.17 /
  0.21 \ \ & 0.21 / 0.31 \ \ & 0.28 / 0.36\\ \hline
\end{tabular}
\end{threeparttable}
\end{center}
\label{tab:simu}
\end{table}

\begin{table}[htb]
\centering
\caption{{\bf Performance comparison of Jordan Center method and Time-reversal backward spreading method.} All the nodes are ranked based on Jordan centrality in descending order and reversal time variance in ascending order respectively. The ranking of the source of spreading process on ER, WS and BA networks are averaged over 100 independent realizations. The time delays of links are assumed to follow Gaussian distributions with mean value 1.0 and variance 0.25 and uniform distributions in the range $(0.5,1.5)$, respectively.  The fraction of observers is 0.05. The network size $N$ is 1000 and the average node degree $\langle k\rangle =8$. The mean ranking of source node and its standard deviation are presented.}
\begin{center}
\begin{threeparttable}
\begin{tabular}{ccccc}
\hline\hline
& & ER &  WS & BA\\
&&\multicolumn{3}{c}{(Mean$\pm$Std)}\\
\hline
\multirow{2}{*}{Gaussian} &TRBS & 1.01$\pm$0.10 & 1.36$\pm$0.88 & 2.92$\pm$8.26\\
& Jordan center & 501.06$\pm$285.20 & 500.95$\pm$304.15 & 446.35$\pm$278.48\\
\hline
\multirow{2}{*}{Uniform} &TRBS & 1.08$\pm$0.36 & 1.59$\pm$1.02 & 6.11$\pm$14.48\\
&Jordan center & 491.75$\pm$309.80 & 478.51$\pm$290.18 & 520.63$\pm$317.78 \\
\hline
\end{tabular}
\end{threeparttable}
\end{center}
\label{tab:jordan}
\end{table}

\end{document}